\documentclass{cjaa}                    

\usepackage{graphicx}                   
\input{epsf.sty}                        
\input{psfig.sty}                       

\usepackage{natbib}

\begin{document}

   \title{Observations of VHE $\gamma$-Ray Sources with the MAGIC Telescope
}


   \author{H. Bartko
      \inst{}\mailto{}
      for the MAGIC Collaboration
      }
   \offprints{H. Bartko}                   

   \institute{ Max-Planck-Institute for Physics, Munich, Germany \\
             \email{hbartko@mppmu.mpg.de}
        }

   \date{Received~~2007 month day; accepted~~2007~~month day}

   \abstract{The MAGIC telescope with its 17m diameter mirror is today 
the largest operating single-dish Imaging Air Cherenkov Telescope (IACT).
It is located on the Canary Island La Palma, at an altitude of
2200m above sea level, as part of the Roque de los
Muchachos European Northern Observatory. The MAGIC telescope detects celestial very high energy $\gamma$-radiation in the energy band between about 50~GeV and 10~TeV. Since Autumn of 2004 MAGIC has been taking data 
routinely, observing various
objects like supernova remnants (SNRs), $\gamma$-ray binaries, Pulsars, Active Galactic Nuclei (AGN) and Gamma-ray Bursts (GRB).
We briefly describe the observational strategy, the procedure implemented for the data analysis, and discuss the results for individual sources. An outlook to the construction of the second MAGIC telescope is given.
   \keywords{TeV $\gamma$-ray astrophysics -- super nova remnants, pulsars, binary systems -- AGN, blazars -- EBL -- GRBs -- dark matter}
}

   \authorrunning{H. Bartko for the MAGIC collaboration }            
   \titlerunning{Observations of VHE $\gamma$-Ray Sources with the MAGIC Telescope} 


   \maketitle
%
%
\section{Introduction}           
\label{sect:intro}



One of the most important 'messengers' of many high energy phenomena in our universe are $\gamma$-rays. The detection of very high energy (VHE, $E_{\gamma}>100$~GeV) cosmic $\gamma$-radiation by ground-based Cherenkov telescopes has opened a new window to the Universe, called VHE $\gamma$-ray astronomy. It is a rapidly expanding field with a wealth of new results, particularly during the last two years, due to the high sensitivity of a new generation of instruments. The major scientific objective of $\gamma$-ray astronomy is the understanding of the production, acceleration, transport and reaction mechanisms of VHE particles in astronomical objects. This is tightly linked to the search for sources of the cosmic rays. 
The MAGIC (Major Atmospheric $\gamma$-ray Imaging Cherenkov) telescope is one of the new generation of Imaging Air Cherenkov Telescopes (IACT) for VHE $\gamma$-ray astronomy. 

The physics program of the MAGIC telescope includes topics, 
both of fundamental physics and astrophysics. 
This article is structured as follows: in section \ref{sec:intro} the MAGIC telescope is presented and the data analysis is explained. The main part of this paper reviews the results of observations with the MAGIC telescope: in section \ref{sec:galactic} the results of Galactic sources and in section \ref{sec:extra_gal} the results of extragalactic sources are described. Sections \ref{sec:GRB} and \ref{sec:DM} deal with the search for $\gamma$-ray emission from Gamma Ray Bursts (GRBs) and dark matter particle annihilation. 
Finally, section \ref{sec:conclusion} contains the conclusions and an outlook to the second MAGIC telescope.

\section{The MAGIC Telescope}  \label{sec:intro}

MAGIC \citep{MAGIC-commissioning,CortinaICRC} 
is currently the largest single-dish Imaging
Air Cherenkov Telescope (IACT) in operation. Located on the Canary
Island La Palma ($28.8^\circ$N, $17.8^\circ$W, 2200~m a.s.l.), it 
has a 17-m diameter tessellated parabolic mirror,
supported by a light-weight carbon fiber frame. It is equipped
with a high-quantum-efficiency 576-pixel $3.5^\circ$ field-of-view photomultiplier tube (PMT)
camera. The analog signals are transported via optical fibers to
the trigger electronics and were read-out by a 300 MSampels/s FADC system till March 2007 and by a new 2 GSamples/s FADC system \citep{MUX_tests} thereafter. 



The MAGIC telescope can operate under moderate moonlight or twilight conditions \citep{MAGIC_moon}. For these conditions, no change in the high voltage settings is necessary as the camera PMTs were especially designed to avoid high currents.



The data analysis is generally carried out using the standard MAGIC
analysis and reconstruction software \citep{Magic-software}, the
first step of which involves the FADC signal reconstruction and the 
calibration of the raw data
\citep{MAGIC_calibration,MAGIC_signal_reco}. After calibration, image-cleaning
tail cuts are applied \citep[see e.g.][]{Fegan1997}.
The camera images are parameterized by
image parameters \citep{Hillas_parameters}. The
Random Forest method (see \citet{RF,Breiman2001} for a detailed
description) was applied for the $\gamma$/hadron separation 
 \citep[for a review see e.g.][]{Fegan1997} 
 and the energy estimation.



For each event, the arrival direction of the primary $\gamma$-ray
candidate in sky coordinates is estimated using the DISP-method resulting in VHE $\gamma$-ray sky maps
\citep{wobble,Lessard2001,MAGIC_disp}. The angular resolution of this procedure is $\sim 0.1^\circ$, while the source localization in the sky is provided with a systematic error of $1'$ \citep{MAGIC_Crab}.

The differential
VHE $\gamma$-ray spectrum
($\mathrm{dN}_{\gamma}/(\mathrm{dE}_{\gamma} \mathrm{dA}
\mathrm{dt})$ vs. true $\mathrm{E}_{\gamma}$) is 
corrected (unfolded) for the instrumental energy resolution
\citep{MAGIC_unfolding}.
All fits to the spectral points take into account the correlations between the spectral points that are introduced by the unfolding procedure.
%
%
%
%
The systematic error in the flux level determination depends on the slope of the $\gamma$-ray spectrum. It is
typically estimated to be 35\% and the systematic error in the
the spectral index is 0.2 \citep{MAGIC_GC,MAGIC_Crab}.


\section{Galactic Sources} \label{sec:galactic}

The observations with the MAGIC telescope 
included the following
types of objects: VHE $\gamma$-ray sources coincident with supernova remnants (section \ref{sec:snr}), the Galactic Center (section \ref{sec:gc}), $\gamma$-ray binaries (section \ref{sec:lsi}), pulsars and pulsar wind nebulae (section \ref{sec:crab}).


\vspace{-0.3cm}
\subsection{Unidentified VHE $\gamma$-ray sources coincident with Supernova remnants}
\label{sec:snr}


Shocks produced by supernova explosions are assumed to be the source
of the galactic component of the cosmic ray flux~\citep{zwicky}. 
In inelastic collisions of high energy cosmic rays with ambient matter $\gamma$-rays and neutrinos are produced. These neutral particles give direct information about their source, as their trajectories are not affected by the Galactic and extra Galactic magnetic fields in contrast to the charged cosmic rays. However, not all VHE $\gamma$-rays from galactic sources are due to the interactions of cosmic rays with ambient matter. There are also other mechanisms for the production of VHE $\gamma$-rays like the inverse Compton up-scattering of ambient low energy photons by VHE electrons. For each individual source of VHE $\gamma$-rays, the physical processes of particle acceleration and $\gamma$-ray emission in this source have to be determined. A powerful tool is the modeling of the multiwavelength emission of the source taking into account the ambient gas density as traced by CO observations \citep{Torres2002}.

Within its program of observation of galactic sources, MAGIC has
taken data on a number of supernova remnants, resulting in the discovery of VHE $\gamma$-ray emission from a source in the SNR IC443, MAGIC J0616+225 \citep{MAGIC_IC443}. Moreover, two recently discovered VHE $\gamma$-ray sources, which are spatially coincident with SNRs, HESS~J1813-178 and HESS~J1834-087 \citep{Aharonian2005b} have been observed with the MAGIC telescope \citep{MAGIC_1813,MAGIC_1834}. Recently, also VHE $\gamma$-rays have been observed from the SNR \textbf{Cas A} \citep{MAGIC_CasA}.

\begin{figure}[h]
  \begin{minipage}[t]{0.5\linewidth}
  \centering
  \includegraphics[height=60mm]{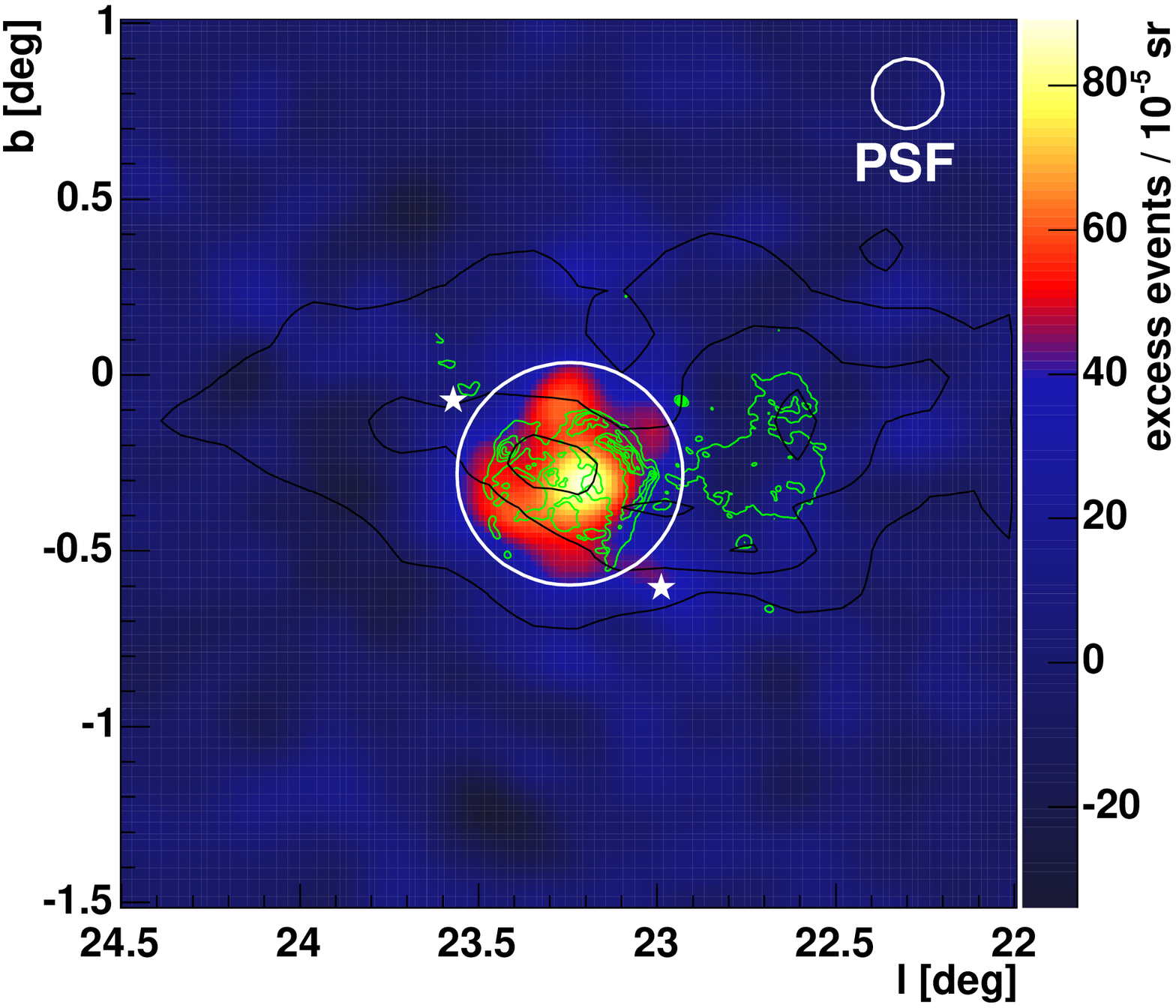}
  \vspace{-5mm}
  \caption{{\small Sky map of $\gamma$-ray candidate events (background
subtracted) in the direction of HESS J1834-087 for an energy threshold
of about 250~GeV. The source is clearly extended with respect to
the MAGIC PSF (small white circle). The two white stars denote the tracking positions of the MAGIC telescope. Overlayed are $^{12}$CO  emission contours
(black) from \citet{Dame2001} and contours of 90 cm VLA radio data
from \citet{White2005} (green). The $^{12}$CO contours are at
25/50/75 K km/s, integrated from 70 to 85 km/s in velocity, the
range that best defines the molecular cloud associated with W41.
The contours of the radio emission are at
0.04/0.19/0.34/0.49/0.64/0.79 Jy/beam, chosen for best showing
both SNRs G22.7-0.2 and G23.3-0.3 at the same time. Clearly, there
is no superposition with SNR G22.7-0.2. The central white circle
denotes the source region integrated for the spectral analysis. \citep{MAGIC_1834}.} \label{fig:sky_W41} }
  \end{minipage}%
  \begin{minipage}[t]{0.5\textwidth}
  \centering
  \includegraphics[height=60mm]{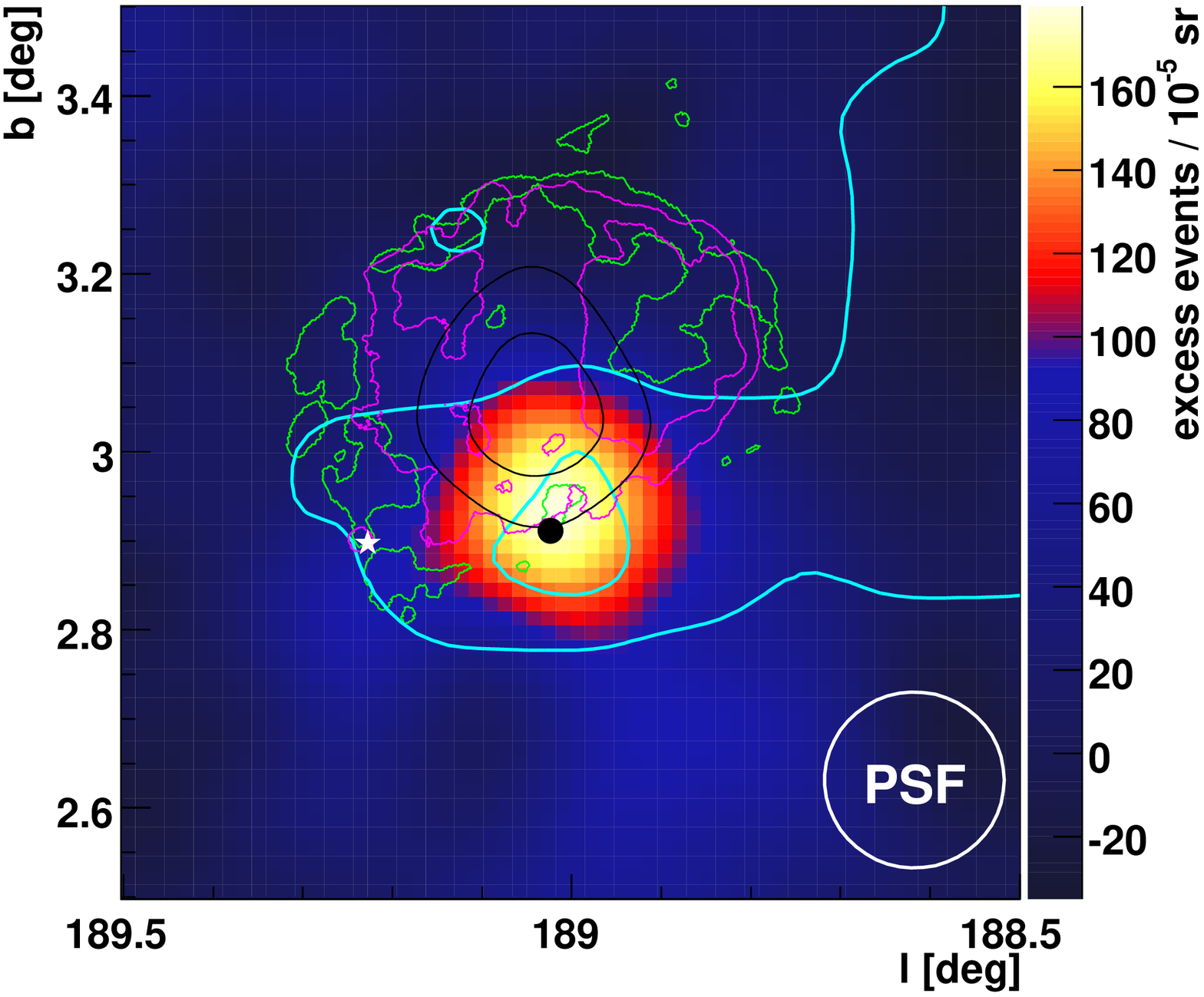}
  \vspace{-5mm}
  \caption{{\small Sky map of $\gamma$-ray candidate events (background
subtracted) in the direction of MAGIC~J0616+225 for an energy threshold
of about 150~GeV. The cyan $^{12}$CO contours \citep{Dame2001} are at
7 and 14 K km/s, integrated from -20 to 20 km/s in velocity, the
range that best defines the molecular cloud associated with IC~443.
The  green contours of 20 cm VLA radio data
\citep{Condon1998} are at
5 mJy/beam, chosen for best showing
both the SNR IC~443. 
The purple Rosat X-ray contours \citep{AsaokaAschenbach1994} are at 700 and 1200 counts / $6 \cdot 10^{-7}$ sr. The black EGRET contours \citep{Hartman1999} represent a 68\% and 95\% statistical probability that a single source lies within the given contour.
The white star denotes the position of the pulsar CXOU J061705.3+222127 \citep{Olbert2001}. The black dot shows the position of the 1720 MHz OH maser \citep{Claussen1997}. The white circle shows the MAGIC PSF of $\sigma = 0.1^{\circ}$. \citep{MAGIC_IC443}.}  \label{fig:sky_IC443}}
  \end{minipage}%

\end{figure}

\textbf{IC443} is a well-studied shell-type SNR near the Galactic Plane with a diameter of 45' at a distance of about 1.5~kpc. It is a prominent source and it has been studied from radio waves to $\gamma$-rays of energies around 1~GeV. \citet{Gaisser1998} as well as other groups extrapolated the energy spectrum of 3EG~J0617+2238 into the VHE $\gamma$-ray range and predicted readily observable fluxes. Nevertheless, previous generation IACTs have only reported upper limits to the VHE $\gamma$-ray emission \citep{Khelifi2003,Holder2005}.
The observation of IC~443 using the MAGIC Telescope has
led to the discovery of a new source of VHE $\gamma$-rays, MAGIC~J0616+225. The flux level of MAGIC~J0616+225 is lower and the energy spectrum (fitted with a power law of slope $\Gamma=-3.1\pm 0.3$) is softer than the predictions \citep{Gaisser1998}.
The coincidence
of the VHE $\gamma$-ray source with SNR IC~443 suggests this
SNR as a natural counterpart. 
A massive molecular cloud and OH maser emissions are located at
the same sky position as that of MAGIC~J0616+225, see figure \ref{fig:sky_IC443}. This suggests that a hadronic origin of the VHE $\gamma$-rays is possible. 
However, other mechanisms for the VHE $\gamma$-ray emission cannot be excluded yet.

\textbf{HESS~J1834-087} is spatially coincident with the SNR G23.3-0.3 (W41). W41 is an asymmetric shell-type SNR, with a diameter of 27' at a distance of $\sim 5$~kpc. It is a prominent radio source, and only recently \citet{Landi2006} found a faint X-ray source within the area of W41 in data from the Swift satellite and \citep{ti06} found an extended X-ray feature spatially coincident with the VHE $\gamma$-ray emission. As in the case of IC~443, the VHE $\gamma$-radiation of W41 is associated with a large molecular complex called "[23,78]''~\citep{da86}, see figure \ref{fig:sky_W41}. Although the mechanism responsible
for the VHE $\gamma$-radiation has not yet been clearly identified,
it could be produced by high energy hadrons interacting with the
molecular cloud.


\textbf{HESS~J1813-178} is spatially coincident with SNR G12.8-0.0 with a diameter of 2' at a distance of $\sim 4$~kpc. It exhibits relatively faint radio and X-radiation. This source is also located in a relatively high-density environment \citep{Lemiere2005}. Recently, inside the SNR shell a putative pulsar wind nebula was discovered \citep{he07,Funk2007}.


These results confirm that Galactic VHE $\gamma$-ray sources are usually spatially correlated with SNRs. Nevertheless, the exact nature of the parent particles of the VHE $\gamma$-rays, their acceleration (in SNR shocks or PWN), and the processes of $\gamma$-ray emission need (or still require) further study.

\subsection{The Galactic Center}
\label{sec:gc}

The Galactic Center region contains many remarkable objects which may be responsible for high-energy processes generating $\gamma$-rays: A super-massive black hole, supernova remnants, candidate pulsar wind nebulae, a high density of cosmic rays, hot gas and large magnetic fields. Moreover, the Galactic Center may appear as the brightest VHE $\gamma$-ray source from the annihilation of possible dark matter particles \citep{DM_ICRC} of all proposed dark matter particle annihilation sources. 


The Galactic Center was observed with the MAGIC telescope~\citep{MAGIC_GC} under large zenith angles, resulting in the measurement of a differential $\gamma$-ray flux, consistent with a steady, hard-slope power law between 500~GeV and about 20~TeV, with a spectral index of $\Gamma=-2.2\pm 0.2$.
This result confirms the previous measurements by the HESS collaboration. The VHE $\gamma$-ray emission does not show any significant time variability; the MAGIC measurements rather affirm a steady emission of $\gamma$-rays from the GC region on time scales of up to one year.

The VHE $\gamma$-ray source is centered at (RA, Dec)=(17$^{\mathrm{h}}45^{\mathrm{m}}20^{\mathrm{s}}$, -29$^\circ2'$). The excess is point-like, its location is consistent with SgrA$^*$, the candidate PWN G359.95-0.04 as well as SgrA East.
The nature of the source of the VHE $\gamma$-rays has not yet been (or yet to be) identified. The power law spectrum up to about 20~TeV disfavours dark matter annihilation as the main origin of the detected flux, see also \citet{HESS_GC_PRL}.
The absence of flux variation indicates that the VHE $\gamma$-rays are rather produced in a steady object such as a SNR or a PWN, and not in the central black hole.



\subsection{The $\gamma$-ray binaries} 
\label{sec:lsi}

\begin{figure}[!ht]
\centering
\includegraphics[width=\textwidth]{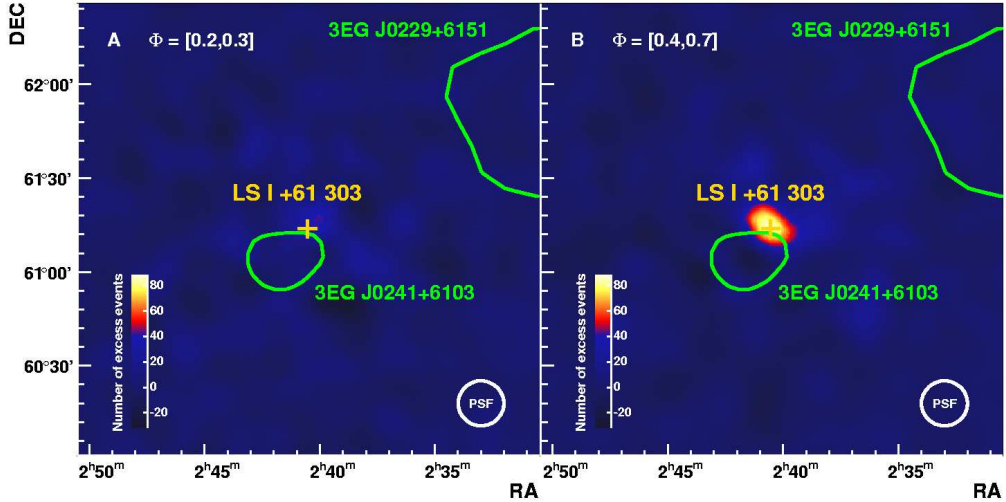}
\caption{Smoothed maps of $\gamma$-ray excess events above 400~GeV around LS~I~+61~303. (A) 15.5 hours corresponding to data around periastron, i.e. between orbital phases 0.2 and 0.3. (B) 10.7 hours at orbital phase between 0.4 and 0.7. The number of events is normalized in both cases to 10.7 hours of observation. The position of the optical source LSI +61 303 (yellow cross) and the 95\% confidence level contours for 3EG J0229+6151 and 3EG J0241+6103 (green contours) \citep{Hartman1999}, are also shown. The bottom-right circle shows the size of the point spread function of MAGIC (1$\sigma$ radius). No significant excess in the number of $\gamma$-ray events is detected around periastron passage, while it shows up clearly (9.4$\sigma$ statistical significance) at later orbital phases, in the location of LS~I~+61~303. \citep{MAGIC_LSI}.
}
\label{fig:lsi-skymap}
\vspace{-0.5cm}
\end{figure}

The $\gamma$-ray binary system \textbf{LS~I~+61~303} is composed of a B0 main sequence star
with a circumstellar disc, i.e. a Be star, located at a distance of
$\sim$2 kpc. A compact object of unknown nature (neutron star or black
hole) is orbiting around it, in a highly eccentric ($e=0.72\pm0.15$)
orbit.  
LS~I~+61~303 was observed with MAGIC for 54 hours
between October 2005 and March 2006~\citep{MAGIC_LSI}. 
The reconstructed $\gamma$-ray sky map is shown in
figure~\ref{fig:lsi-skymap}. The data were first divided into 
two different samples, around periastron passage (0.2-0.3)
and at higher (0.4-0.7) orbital phases. No significant excess in the
number of $\gamma$-ray events is detected around periastron passage,
whereas there is a clear detection (9.4$\sigma$ statistical significance) at
later orbital phases.
Two different scenarios were discussed to explain this high energy
emissions: the microquasar scenario where the $\gamma$-rays are produced
in a radio-emitting jet; or the pulsar binary scenario, where 
they are produced in the shock which is generated by the interaction
of a pulsar wind and the wind of the massive companion.

Recently, an excess $4.1 \sigma$ significance (after trial correction) of $\gamma$-ray candidate events over the expected background was observed during 79 minutes of one night for the black hole X-ray binary (BHXB) \textbf{Cygnus X-1} \citep{MAGIC_CygX1}. Moreover, VHE $\gamma$-ray emission has been observed from the high mass X-ray Binary LS~5039 by the H.E.S.S. collaboration \citep{HESS_5039}.

\subsection{Pulsars and Pulsar Wind Nebulae} \label{sec:crab}

The Crab Nebula is a bright and steady emitter of GeV and TeV energies, 
and is therefore an excellent calibration candle. This object has
been observed intensively in the past, over a wide range of
wavelengths. 

The energy domain between 10 and 100~GeV is of particular interest, as both
the Inverse Compton peak of the spectral energy distribution 
and the cut-off of the pulsed emission is expected in this energy range.

\begin{figure}[h]
  \begin{minipage}[t]{0.5\linewidth}
  \centering
  \includegraphics[height=50mm]{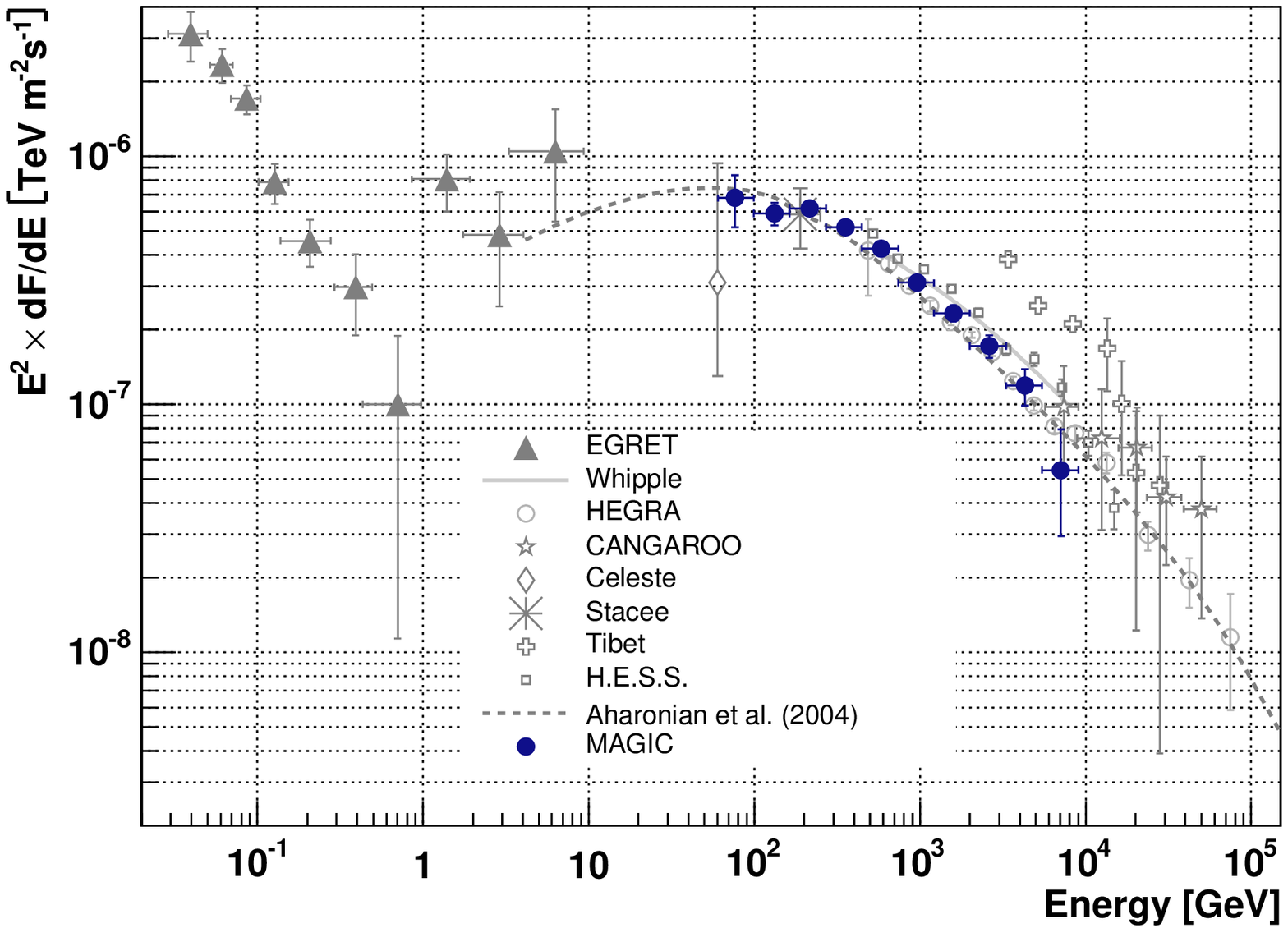}
  \vspace{-5mm}
  \caption{{\small Spectral energy distribution of the  $\gamma$-ray emission of the Crab Nebula. The measurements below 10~GeV are from the EGRET, the measurements above are from ground-based experiments. Above 400~GeV the MAGIC data are in agreement with measurements of other IACTs. The dashed line represents a model prediction by \citet{Aharonian_Crab}. \citep{MAGIC_Crab}.}  \label{fig:crab}}
  \end{minipage}%
  \begin{minipage}[t]{0.5\textwidth}
  \centering
  \includegraphics[height=50mm]{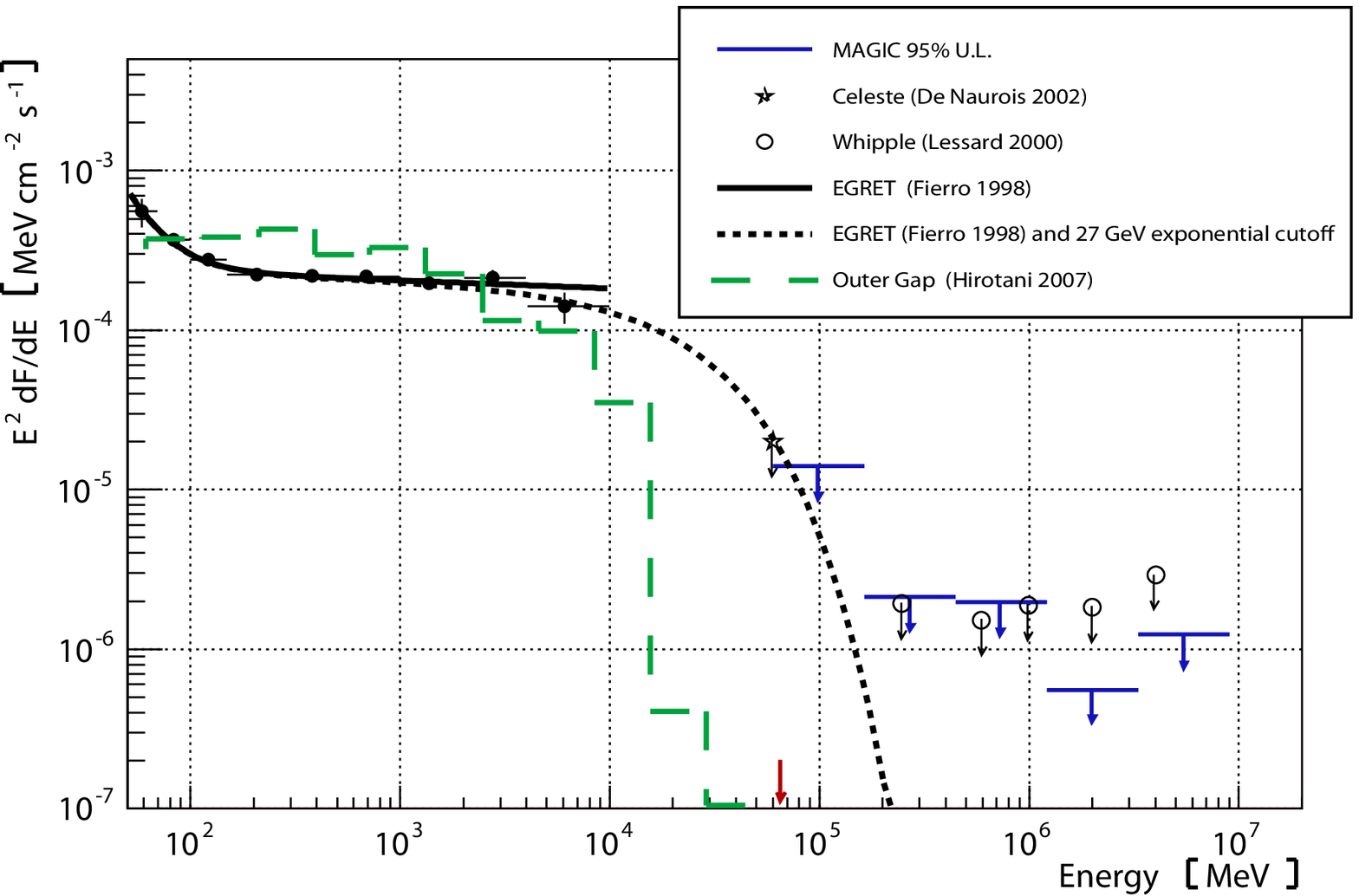}
  \vspace{-5mm}
  \caption{{\small Upper limits (95\% CL.) on the pulsed $\gamma$-ray flux from the Crab
                        Pulsar; upper limits in bins of energy are given
                        by the blue points. The upper limit on the
                        cutoff energy of the pulsed emission is
                        indicated by the dashed line. The
                        analysis threshold to derive the upper
                        limit on the cutoff energy is indicated by the red arrow. \citep{MAGIC_Crab}.
                        }  \label{fig:crab_pulsar}}
  \end{minipage}%
\end{figure}

A significant amount of MAGIC's observation time has been devoted to observing the Crab Nebula, both for
technical (because it is a strong and steady emitter) and astrophysical studies. A sample of
16~hours of selected data has been used to measure
the energy spectrum between 60~GeV and 9~TeV, and the result is shown in
figure~\ref{fig:crab}~\citep{MAGIC_Crab}. Also, a search
for pulsed $\gamma$-ray emission from the Crab Pulsar has been carried out. Figure~\ref{fig:crab_pulsar} shows the derived (95\% CL.) upper limits.



Pulsed $\gamma$-ray emission was also searched for from the pulsar PSR B1951+32. 
A 95\% CL. of $4.3 \cdot 10^{-11}$~cm$^-2$~sec$^{-1}$ was obtained for the flux of pulsed $\gamma$-ray emission for $E_{\gamma}>75$~GeV and of $1.5 \cdot 10^{-11}$~cm$^-2$~sec$^{-1}$ for the steady emission for $E_{\gamma}>140$~GeV \citep{MAGIC_1951}.

\section{Extragalactic Sources} \label{sec:extra_gal}

The detection and characterization of VHE $\gamma$-ray emitting Active Galactic Nuclei (AGN) is one of the main goals of ground--based $\gamma$-ray astronomy. 
The observational results can be used to explore the physics of the relativistic jets in AGNs, to understand the origin of the VHE $\gamma$-rays as well as to correlate with each other the fluxes of photons in different energy bands (optic, X-rays and $\gamma$-rays). Moreover, perform population studies of AGNs can be performed, information about the extragalactic background light (EBL) density can be extracted and even questions and even the a possible vacuum refractive index, that might be induced by quantum gravity, can be probed.
%
%
%
The set of extragalactic objects which were observed by the MAGIC telescope comprises known TeV-emitting blazars (AGNs with a jet axis close to the line of sight) as well as VHE $\gamma$-ray candidate sources
such as selected high- and low-frequency peaked BL Lacs (HBLs and LBLs). Moreover, other non--blazar objects like the ULIRG \mbox{Arp 220} have also been observed, although none of these observations resulted in a positive detection so far \citep{MAGIC_Arp220}.

In section \ref{sec:source_candidates} the discoveries of VHE $\gamma$-rays from candidate sources will be described, while in section \ref{sec:known_AGN} the observation of known TeV blazars is reviewed. 



\subsection{Discoveries of VHE $\gamma$-Rays from Candidate Sources} \label{sec:source_candidates}

The selection of candidates for VHE $\gamma$-ray emitting sources follows criteria based on the spectral properties of the considered objects at lower frequencies, see e.g. \citet{MAGIC_AGN_search}. Using both Synchrotron Self--Compton (SSC) and hadronic models, the spectral energy distribution of the candidate AGN is extrapolated to MAGIC energies to predict its observability. The preferred candidates are usually strong X-ray emitters, but selections based on the optical band have also been followed. 


\textbf{1ES 1218+30.4} \citep{MAGIC_1218} is the first source discovered by MAGIC and one of the most distant VHE $\gamma$-ray sources known so far. This HBL, which has a redshift of $z=0.182$, was previously observed by Whipple and HEGRA, but only upper flux limits were determined. MAGIC observed it during 8.2~h in January 2005, obtaining a $\gamma$-ray signal of 6.4~$\sigma$ significance in the 87 to 630 GeV energy range. The differential energy spectrum can be fitted by a simple power law with a photon index of $3.0\pm0.4$. No time variability on timescales of days was found within statistical errors.


\begin{figure}[h]
  \begin{minipage}[t]{0.5\linewidth}
  \centering
  \includegraphics[height=50mm]{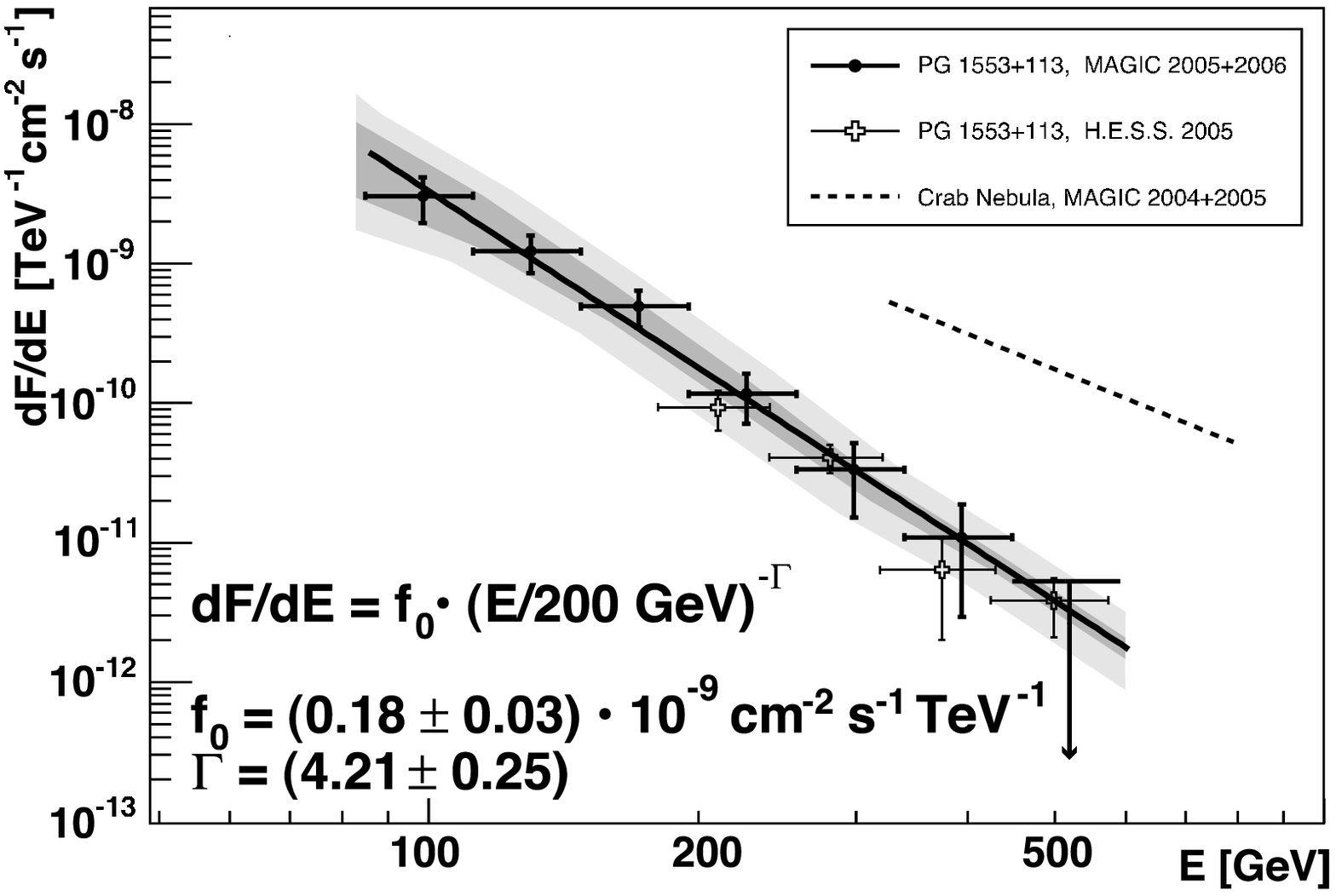}
  \vspace{-5mm}
  \caption{{\small Differential energy spectrum of PG 1553+113 as derived from
the combined 2005 and 2006 data. The MAGIC Crab Nebula energy spectrum and
the H.E.S.S. PG 1553+113 energy spectrum \citep{hess1553} have been included for
comparison. \citep{MAGIC_1553}.}   \label{fig:1553}}
  \end{minipage}%
  \begin{minipage}[t]{0.5\textwidth}
  \centering
  \includegraphics[height=50mm]{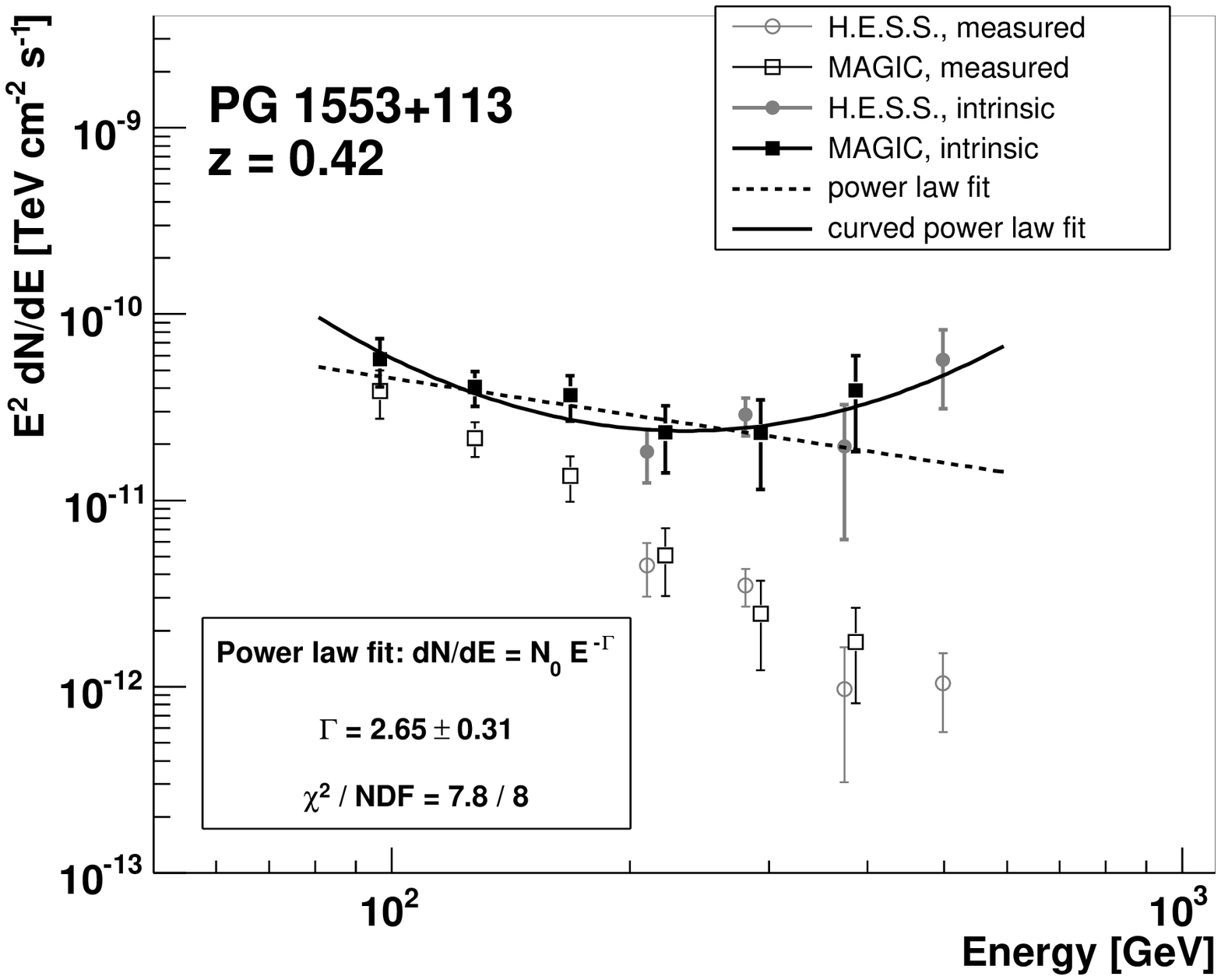}
  \vspace{-5mm}
  \caption{{\small Observed differential energy spectrum of PG 1553+113 
         multiplied by E$^2$ to represent the spectral energy density by H.E.S.S. and MAGIC (open symbols) and source intrinsic spectra (full symbols), corrected for the attenuation by the EBL assuming $z=0.42$. 
        %
	\citep{Mazin2007}}  \label{fig:1553_mazin}}
  \end{minipage}%
\end{figure}

\textbf{PG 1553+113}  is a distant BL Lac of undetermined redshift. It was recently detected by the H.E.S.S. collaboration \citep{hess1553} as well as by the MAGIC collaboration \citep{MAGIC_1553}. A VHE $\gamma$-ray signal has been observed by the MAGIC telescope with an overall significance of 8.8~$\sigma$, showing no significant flux variations on a daily timescale. However, the flux observed in 2005 was significantly higher compared to 2006. The MAGIC measurements reach to substantially lower energies than the HESS
measurements do. The differential energy spectrum between 90 and 500~GeV can be well described by a power law with photon index of $4.2\pm0.3$, being steeper than that of any other known BL Lac object, see figure \ref{fig:1553}. This spectrum can be used to derive an upper limit on the source redshift. Assuming an EBL model by \citet{EBLKneiske} and a limit of $\alpha_{\mathrm{int}}<-1.5$ for photon index $\alpha_{\mathrm{int}}$ of the intrinsic source spectrum \citep{limit}, an upper limit on the source redshift of $z<0.78$ has been derived. \citet{Mazin2007} find that a redshift above $z = 0.42$ implies a possible break of the intrinsic spectrum at about 200 GeV. Assuming that such a break is absent, they obtain a much stronger upper limit of $z < 0.42$, see figure \ref{fig:1553_mazin}.  




\textbf{Mkn 180} \citep{MAGIC_Mrk180} is an HBL ($z = 0.045$) that had an optical outburst in March 2006 observed by the KVA 35~cm telescope, which is also located at the Roque de los Muchachos Observatory. The optical outburst triggered the observations with the MAGIC telescope in the GeV--TeV band, resulting in the first detection of VHE $\gamma$-ray emission from this source. A total of 12.4~h of data were recorded during eight nights, giving a $5.5 \sigma$ significance detection. The integral flux above 200~GeV corresponded to 11\% of the Crab Nebula flux, and the differential energy spectrum could be fitted by a power law (photon index $3.3\pm0.7$).

\textbf{Bl Lacertae} \citep{MAGIC_BlLac} is a low-frequency peaked BL Lac (LBL) object at $z=0.069$. It was observed with the MAGIC telescope from August to December 2005 (22.2 hrs), and from July to September 2006 (26.0 hrs). A very high energy (VHE) $\gamma$-ray signal was discovered with a 5.1 sigma excess in the 2005 data. Above 200 GeV, an integral flux of 3\% of the Crab Nebula flux was measured.
The differential energy spectrum between 150 and 900 GeV is rather steep, with a photon index of $-3.6 \pm 0.5$. For the first time, a clear detection of VHE $\gamma$-ray emission from an LBL object was obtained. During the observation, the light curve showed no large flux variation. The 2006 data show no significant excess. 

Recently, also \textbf{1ES 1011+496} was discovered as a source of VHE $\gamma$-rays by the MAGIC telescope \cite{MAGIC_1011}.
 

\subsection{Observation of known TeV blazars} \label{sec:known_AGN}

The sensitivity and lower energy threshold of MAGIC as compared to the former generation of $\gamma$-ray telescopes, allows a detailed study of the spectral features and flux variations of known TeV emitters.

\begin{figure}[h!]
  \begin{minipage}[t]{0.5\linewidth}
  \centering
  \includegraphics[height=80mm]{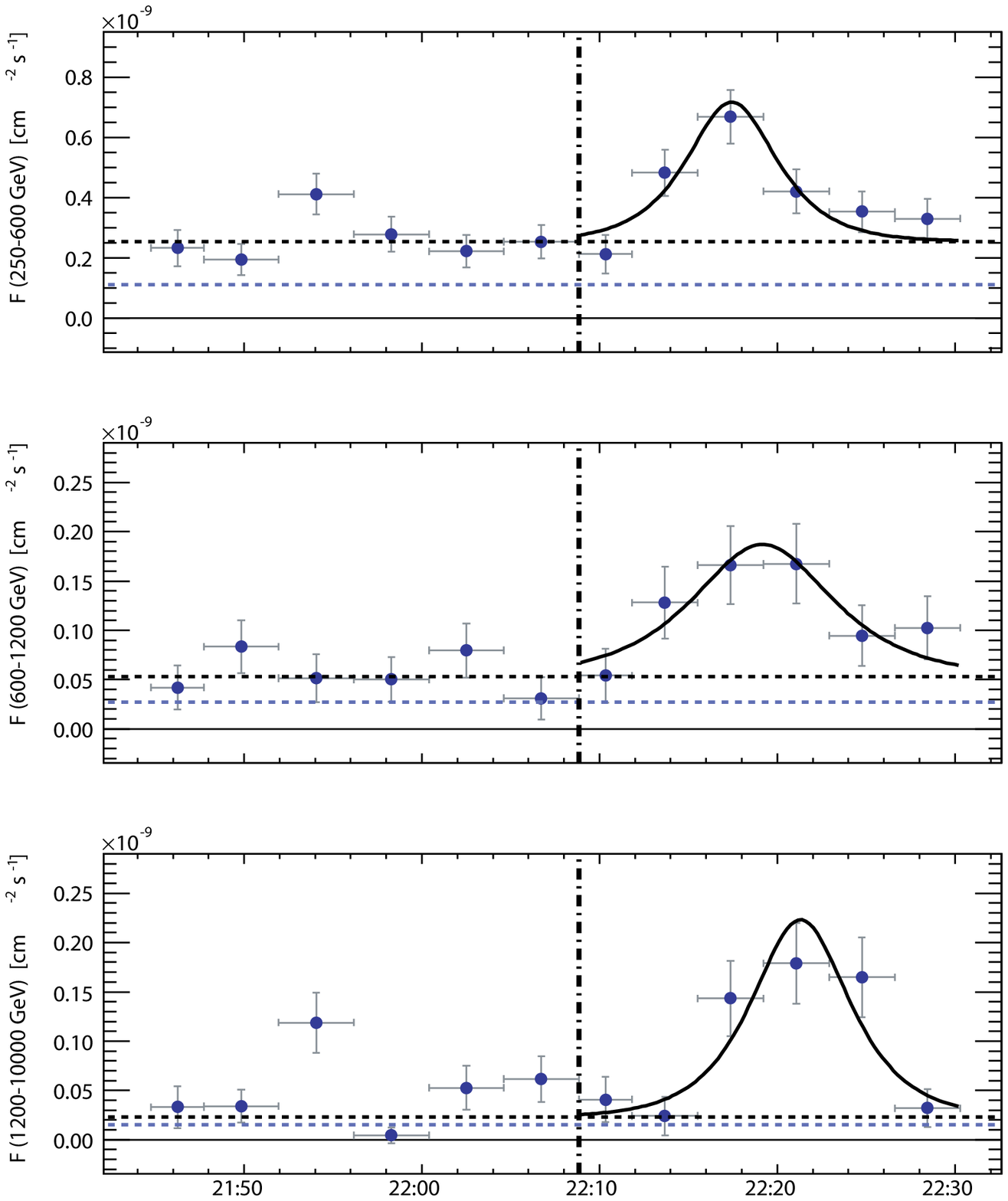}
  \vspace{-5mm}
  \caption{{\small VHE $\gamma$-ray light curve for the night 2005 July 9 with a time binning of 4 minutes, 
and separated in 3 different energy bands, from the top to the bottom, 
0.25-0.6 TeV, 0.6-1.2 TeV, 1.2-10 TeV.
The vertical bars denote 1$\sigma$ statistical 
uncertainties. For comparison, the Crab emission is also shown as a lilac dashed horizontal line. The vertical dot-dashed line divides the data into 'stable' (i.e., pre-burst) and 'variable' (i.e., in-burst) emission. The horizontal black dashed line represents the average of the 'stable' emission. The 'variable' (in-burst) of all energy ranges were fitted with a flare model. \citep{MAGIC_Mrk501}.}  \label{fig:Mrk501_flare}}
  \end{minipage}%
  \begin{minipage}[t]{0.5\textwidth}
  \centering
  \includegraphics[height=70mm]{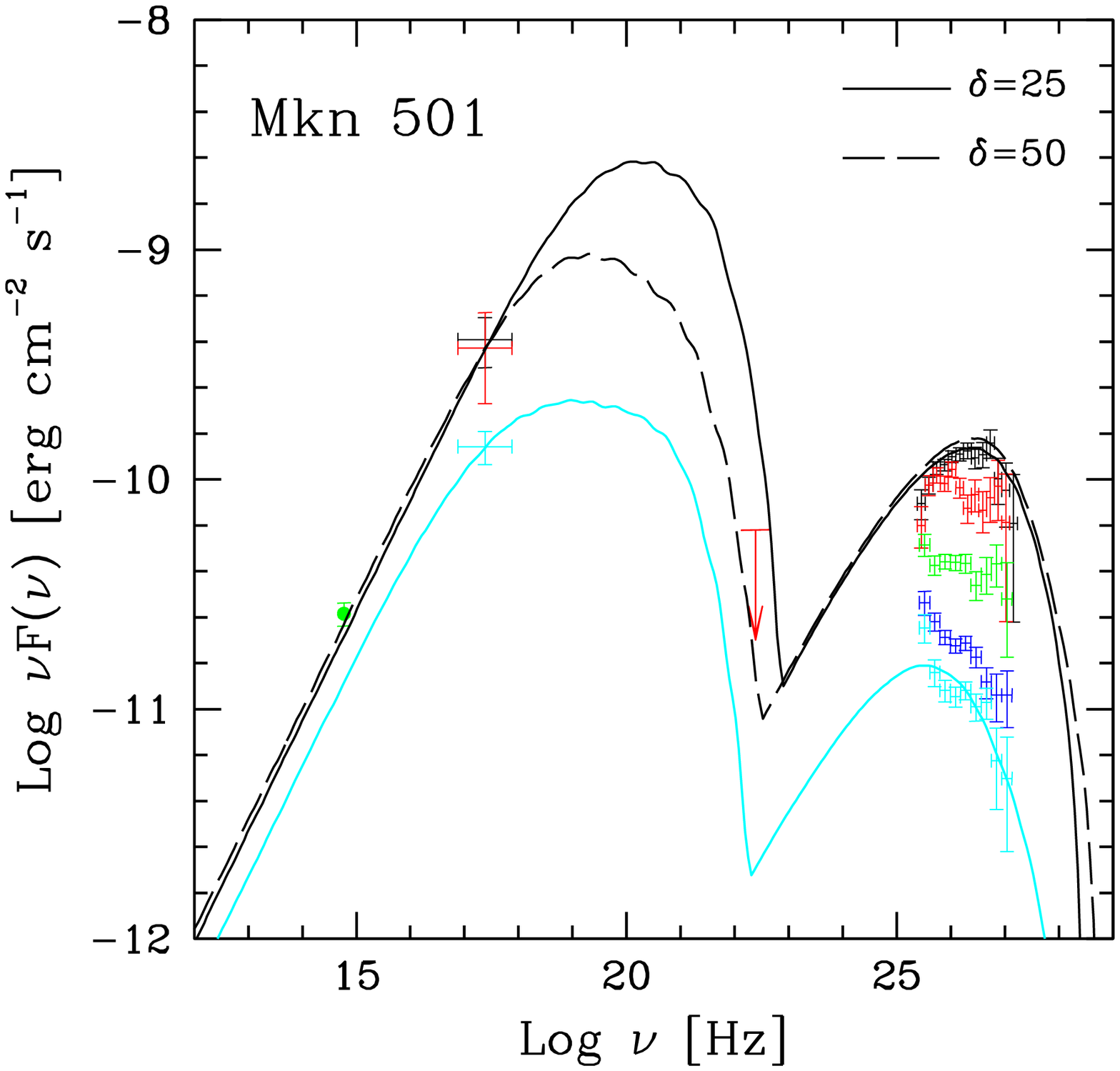}
  \vspace{-5mm}
  \caption{{\small Overall SED from Mrk501. Optical data from the  KVA Telescope: green circle, X-ray data from RXTE ASM for June 30 (black points), for July 9 (red) and  for the other nights combined (light blue). VHE data from MAGIC: black points (June 30), red points (July 9), green points ('high flux' data-set), dark blue  points ('medium flux' data-set), and light blue points ('low flux' data-set). The VHE spectra are corrected for EBL extinction using \citep{EBLKneiske}'s 'Low' EBL model. The highest and the lowest state were fitted with a one zone SSC model.
\citep{MAGIC_Mrk501}.} \label{fig:Mrk501_mwl}}
  \end{minipage}%
 \end{figure}

\textbf{Mkn 421} \citep{MAGIC_Mrk421} is the closest TeV blazar ($z=0.031$) and the first extragalactic VHE source detected with a ground--based $\gamma$-ray telescope \citep{punch}. MAGIC has observed this source between November 2004 and April 2005, obtaining 25.6~h of data. Integral flux variations up to a factor of four are observed between different observation nights, although no significant intra--night variations have been recorded, despite the high sensitivity of the MAGIC telescope for this kind of search. This flux variability shows a clear correlation between $\gamma$-ray and X-ray fluxes, favoring leptonic emission models. The energy spectrum between 100~GeV and 3~TeV shows a clear curvature. After correcting the measured spectrum for the effect of $\gamma$-attenuation caused by the EBL  assuming a model of \citet{primack}, there is an indication of an inverse Compton peak around 100~GeV.

\textbf{1ES 2344+514} ($z = 0.044$) was first detected in 1995 by the Whipple collaboration when it was in a flaring state \citep{w2344}. The HEGRA collaboration reported later an evidence for a signal on a 4 sigma level \citep{h54AGN}. 
MAGIC obtained a VHE $\gamma$-ray signal with 11.0~$\sigma$ significance from 23.1~h of data \citep{MAGIC_2344}, measuring its energy spectrum from 140 GeV to 5 TeV. The source was in the quiescent state during the observations, with a flux level compatible with the HEGRA results, but showing a softer spectrum. 

\textbf{1ES 1959+650} ($z = 0.047$) is a very interesting source, as it showed in 2002 a VHE $\gamma$-ray flare without any counterpart in X-rays \citep{kraw}. This behavior cannot be easily explained by the SSC mechanism in relativistic jets that successfully explains most of the VHE $\gamma$-ray production in other HBLs. MAGIC observed this object during 6~h in 2004, when it was in low activity both in optical and X-ray bands, detecting a $\gamma$-ray signal with 8.2~$\sigma$ significance \citep{MAGIC_1959}. The differential energy spectrum between 180~GeV and 2~TeV can be fitted with a power law of photon index $2.72\pm0.14$, which is consistent with the slightly steeper spectrum seen by HEGRA at higher energies \citep{h1959}, also during periods of low X-ray activity.

\textbf{Mkn 501} \citep{MAGIC_Mrk501} is a close TeV blazar ($z = 0.034$), first detected by the Whipple collaboration \citep{w501}. MAGIC observed it during 55~h in 2005, including 34~h in moderate moonlight conditions. The source was in the low state (30-50\% of the Crab Nebula flux for $E>200$~GeV) during most of the observation time, but showed two episodes of fast and intense flux variability, with doubling times of about 2 minutes, see figure \ref{fig:Mrk501_flare}. The energy spectrum was measured from 100~GeV up to 5~TeV. Changes in the spectral slope with the flux level have been observed for the first time on timescales of about 10 minutes. Figure \ref{fig:Mrk501_mwl} shows the overall SED of Mrk501 for different days as well as 'high', 'medium' and 'low' flux data sets. Recently, the timing of photons observed by the MAGIC gamma-ray telescope during this flare was used to probe a vacuum refractive index, that might be induced by quantum gravity \citep{MAGIC_501_QG}.

\section{Gamma Ray Bursts} \label{sec:GRB}

A wealth of gamma ray bursts (GRBs) have been observed since their first detection more than 
30~years ago. Nevertheless, the processes leading to these extraordinarily energetic 
outbursts are still largely unknown.
EGRET, a satellite $\gamma$-ray telescope, measured $\gamma$-rays up to 18~GeV \citep{EGRET_GRB}, and measuring 
them at higher energies with the MAGIC telescope would substantially help to understand them better. 
Such observations would also give a clue
as to their distance, because high-energy $\gamma$-rays at larger distances are increasingly strongly 
absorbed by the intergalactic infrared background radiation. 
More precise localization, as possible in MAGIC, would also help
in identifying them with known sources. Possible correlations between flux 
variations at different
energies might even allow setting limits for the invariance of 
the speed of light, which some models of
quantum gravity claim to be violated.


MAGIC being built specifically also for GRB observations, by minimizing weight and installing powerful
driving motors, an alert for GRB (as they come from satellite experiments) triggers the re-orientation 
of the telescope and the start of the new observation within a maximum time (depending on the position in the sky)
of 40 seconds. On 13 July 2005, triggered by such an alarm from Swift-BAT, MAGIC succeeded for the first
time to track a GRB during its prompt phase 
\citep{MAGIC_GRB050713a}. No significant radiation at
high energy was seen. Thereafter, nine additional GRBs have been targeted with the MAGIC telescope, but in neither data set any evidence for a $\gamma$-ray signal was found. Upper limits for the flux were derived for all events. For the bursts with measured redshift, the upper limits are compatible with a power law extrapolation, when the intrinsic fluxes are evaluated taking into account the attenuation due to the scattering in the Metagalactic Radiation Field (MRF) \citep{MAGIC_GRBs}.

\section{Search for Dark Matter} \label{sec:DM}

We know today, from measuring gravitational effects, that the visible Universe represents only a 
fraction of the matter in the Universe: some 75\% of all matter cannot be seen, and is called "dark matter". 
Dark matter cannot be made up of the same constituents as visible matter, and must be "non-baryonic".
Some theories predict that, with very low probability, such non-baryonic particles upon
collision can produce VHE $\gamma$-rays. Observing such annihilation products would, of course, be
an epochal discovery for cosmology and astrophysics. MAGIC dedicates some observation time to such searches,
despite the small probabilities involved \citep{DM_ICRC}.



\section{Conclusions and Outlook} \label{sec:conclusion}

Using the MAGIC telescope, seven galactic and nine extra-galactic sources of VHE $\gamma$-radiation have been observed. Nine of these objects have been detected before in VHE $\gamma$-rays. The high sensitivity and the low energy threshold of the MAGIC telescope allowed detailed studies of the spectral features of these sources, as well as the observation of flux variability on short timescales.

The MAGIC collaboration is currently constructing a second telescope on the same site at the Roque de los Muchachos Observatory, which will operate in stereo mode with the MAGIC telescope improving the overall sensitivity \citep{MAGIC2}. The MAGIC II telescope is a clone of the existing one with one main improvement: a fine pixelized camera with a cluster design that will allow an upgrade of the photomultipliers to hybrid photon detectors once this technology is ready to be used. The estimated sensitivity of a system of two MAGIC telescopes is a factor of two better than the present sensitivity of the MAGIC telescope.

\paragraph{Acknowledgements.}
We thank the IAC for the excellent working conditions at the
ORM in La Palma. The support of the
German BMBF and MPG, the Italian INFN, the Spanish CICYT is gratefully
acknowledged. This work was also supported by ETH research grant
TH-34/04-3, and the Polish MNiI grant 1P03D01028.

\label{lastpage}

\end{document}